\documentclass[conference,a4paper]{IEEEtran}
\usepackage[left=1in, right=0.7in, bottom=1.4in, top=0.76in]{geometry}
\IEEEoverridecommandlockouts

\usepackage{cite}
\usepackage{amsmath,amssymb,amsfonts}
\usepackage{algorithmic}
\usepackage{graphicx}

\usepackage{booktabs}  
\usepackage{makecell}  
\usepackage{siunitx}   
\usepackage{multirow}

\usepackage[caption=false]{subfig}

\usepackage{textcomp}
\usepackage{xcolor}
\usepackage{fancyhdr}
\usepackage{url}
\def\BibTeX{{\rm B\kern-.05em{\sc i\kern-.025em b}\kern-.08em
    T\kern-.1667em\lower.7ex\hbox{E}\kern-.125emX}}

\usepackage{array}
\usepackage{acronym}

\acrodef{FEEC}{Faculty of Electrical Engineering and Communication}
\acrodef{BUT}{Brno University of Technology}
\acrodef{CIR}{channel impulse response}
\acrodef{RMS}{root mean square}
\acrodef{DS}{delay spread}
\acrodef{MMW}[MMW]{millimeter wave}
\acrodef{FMCW}[FMCW]{frequency modulated continuous wave}
\acrodef{FSPL}[FSPL]{free space path loss}
\acrodef{FFT}[FFT]{fast fourier transform}
\acrodef{IFFT}[IFFT]{inverse fast fourier transform}
\acrodef{USB}[USB]{upper sideband}
\acrodef{LSB}[LSB]{lower sideband}
\acrodef{IRR}{image rejection ratio}
\acrodef{SNR}{signal to noise ratio}
\acrodef{MPC}{multipath components}
\acrodef{LOS}{line-of-sight}
\acrodef{NLOS}{non-line-of-sight}
\acrodef{GPS}{global positioning system}
\acrodef{TX}{transmitter}
\acrodef{RX}{receiver}
\acrodef{ATT}{attenuator}
\acrodef{G}{gain}
\acrodef{RP}{relative power}
\acrodef{PDP}{power delay profile}
\acrodef{OWGA}{open waveguide antenna}
\acrodef{HPBW}{half power beam width}
\acrodef{AGV}{autonomous ground vehicles}
\acrodef{CDF}{cumulative distribution function}
\acrodef{WRC}{world radiocommunication conference}
\acrodef{CIR}{channel impulse response}
\acrodef{RMS}{root mean square}
\acrodef{DS}{delay spread}
\acrodef{MMW}[MMW]{millimeter wave}
\acrodef{FMCW}[FMCW]{frequency modulated continuous wave}
\acrodef{FSPL}[FSPL]{free space path loss}
\acrodef{FFT}[FFT]{fast fourier transform}
\acrodef{IFFT}[IFFT]{inverse fast fourier transform}
\acrodef{USB}[USB]{upper sideband}
\acrodef{LSB}[LSB]{lower sideband}
\acrodef{IRR}{image rejection ratio}
\acrodef{SNR}{signal to noise ratio}
\acrodef{MPC}{multipath components}
\acrodef{GPS}{global positioning system}
\acrodef{TX}{transmitter}
\acrodef{RX}{receiver}
\acrodef{ATT}{attenuator}
\acrodef{G}{gain}
\acrodef{RP}{relative power}
\acrodef{PDP}{power delay profile}
\acrodef{OWGA}{open-ended waveguide antenna}
\acrodef{HPBW}{half power beam width}
\acrodef{LNA}{low noise amplifier}
\acrodef{PA}{power amplifier}
\acrodef{V2V}{vehicle-to-vehicle}
\acrodef{V2I}{vehicle-to-infrastructure}
\acrodef{H}{heights}
\acrodef{ITU}{international telecommunication union}
\acrodef{ISM}{industrial, scientific, and medical}

\fancypagestyle{firstpage}
{
    \fancyhf{}
    \fancyhead[C]{The paper has been presented at the 2025 35th International Conference Radioelektronika (RADIOELEKTRONIKA),Hnanice, Czech Republic, May 12-14, 2025,\\ \url{https://doi.org/10.1109/RADIOELEKTRONIKA65656.2025.11008402}}
    \fancyfoot[C]{This research was funded in part by the National Science Center (NCN), Poland, grant no. 2021/43/I/ST7/03294 (MubaMilWave). For this purpose of Open Access, the author has applied a CC-BY public copyright license to any Author Accepted Manuscript (AAM) version arising from this submission.}
}

\begin{document}
\makeatletter

 \title{Microdiversity and Vegetation Influence on Forward Scattering at 60\,GHz and 80\,GHz}

\author{\IEEEauthorblockN{Radek Zavorka\IEEEauthorrefmark{1}, Ondrej Zeleny\IEEEauthorrefmark{1}, Jiri Blumenstein\IEEEauthorrefmark{1}, Tomas Mikulasek\IEEEauthorrefmark{1}, Rajeev Shukla\IEEEauthorrefmark{2}, Josef Vychodil\IEEEauthorrefmark{1}, \\Jarosław Wojtuń\IEEEauthorrefmark{3}, Niraj Narayan\IEEEauthorrefmark{2}, Aniruddha Chandra\IEEEauthorrefmark{2}, Jan Kelner\IEEEauthorrefmark{3}, Cezary Ziółkowski\IEEEauthorrefmark{3}, Ales Prokes\IEEEauthorrefmark{1}}
\IEEEauthorblockA{\IEEEauthorrefmark{1}Department of Radio Electronics, Brno University of Technology, Brno, Czech Republic}
\IEEEauthorblockA{\IEEEauthorrefmark{2}National Institute of Technology, Durgapur, India}
\IEEEauthorblockA{\IEEEauthorrefmark{3}Institute of Communications Systems, Faculty of Electronics,\\ Military University of Technology, 00-908 Warsaw, Poland}
e-mail: xzavor03@vutbr.cz}

\maketitle
\thispagestyle{firstpage}

\begin{abstract}
Understanding the impact of vegetation and small-scale antenna movements on signal propagation is important for the design and optimization of high-frequency wireless communication systems. This paper presents an experimental study analyzing signal propagation at 60\,GHz and 80\,GHz in the presence of vegetation, with a focus on forward scattering and microdiversity effects. A controlled measurement campaign was conducted in an indoor environment, where the influence of a potted plant placed in the line-of-sight (LOS) path between the transmitter and receiver was investigated. The study examines the effects of antenna micro-shifts on the channel impulse response (CIR), highlighting variations in received power due to small positional changes of the antennas. The results indicate that the 80\,GHz band exhibits higher sensitivity to micro-movements compared to the 60\,GHz band, leading to greater fluctuations in received power.
\end{abstract}


\section{Introduction}
\label{introduction}
The rapid development of wireless communication networks requires a thorough analysis of how electromagnetic waves propagate in various real-world environments. Urban canyons, such as streets between buildings~\cite{canyon, canyon2}, open spaces, and the presence of vegetation, particularly deciduous trees, significantly impact signal transmission~\cite{vege_brazil, vege_china}. Studies have shown that these factors contribute to increased path loss and signal attenuation, requiring careful consideration in network design. Furthermore, as wireless communication moves towards higher frequency bands, the need for research into \ac{MMW} propagation has become important. In this paper, we will focus on the 60\,GHz and 80\,GHz frequency bands. 

The 60\,GHz frequency band, which provides several gigahertz of bandwidth, has been designated by the \ac{ITU} as part of the \ac{ISM} spectrum, allowing for unlicensed use~\cite{ITU_ISM_60}. In contrast, the E-band encompasses the 71–76\,GHz and 81–86\,GHz frequency ranges, offering a total bandwidth of 10\,GHz, and has been considered a potential candidate for 5G applications at \mbox{\ac{WRC}-15}~\cite{wrc19}. Compared to the 60\,GHz band, electromagnetic waves in the E-band experience lower attenuation due to oxygen absorption during atmospheric propagation~\cite{absorb}.

Research has demonstrated that trees, particularly deciduous trees, can significantly attenuate radio signals due to the high water content in their leaves~\cite{leaf_water} and branches, which absorb and scatter electromagnetic waves. Additionally, foliage and branches contribute to multipath effects by inducing diffuse reflections and scattering, resulting in increased path loss and signal distortion. In~\cite{foliage1}, an analysis of cherry tree attenuation showed a loss of 0.4\,dB/m for both co- and cross-polarized antenna configurations. Similarly,~\cite{chavero} identified a vegetation-induced attenuation factor by comparing the average received power in \ac{LOS} and \ac{NLOS} conditions, finding attenuation values between 19\,dB and 26\,dB at 28\,GHz, depending on the polarization of the transmitted signal.

\subsection{Contribution of this paper}

The main contributions of this paper are as follows:
\begin{itemize}
\item This paper presents a measurement campaign of a static channel in a controlled environment to evaluate the influence of foliage on signal propagation through vegetation.
\item The impact of antenna micro-shifts on the channel impulse response is examined.
\item A comparative analysis of the 60\,GHz and 80\,GHz frequency bands is conducted.

\end{itemize}


\section{Description of measured scenarios}
\label{sec:descriptions_scenario}
\begin{figure}[htbp]
    \centering
    \includegraphics[width=0.4\textwidth]{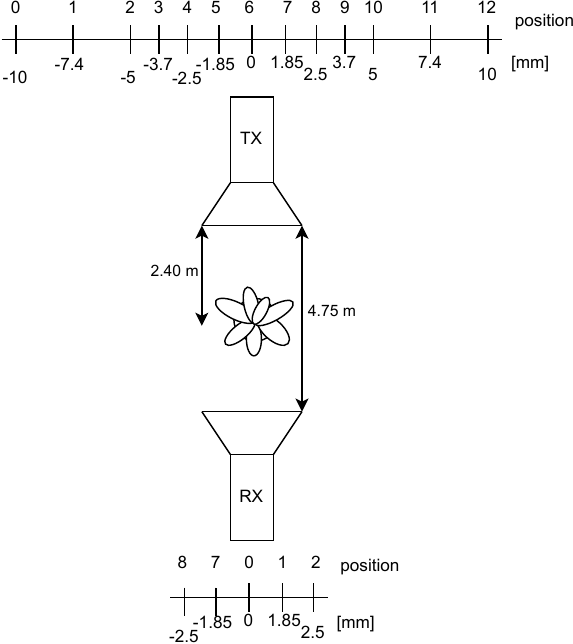}
    \caption{Schematic diagram of the measurement campaign and \ac{TX} and \ac{RX} distances.}
    \label{fig:diagram}
\end{figure}

\begin{figure}[htbp]
    \centering
    \includegraphics[width=0.5\textwidth]{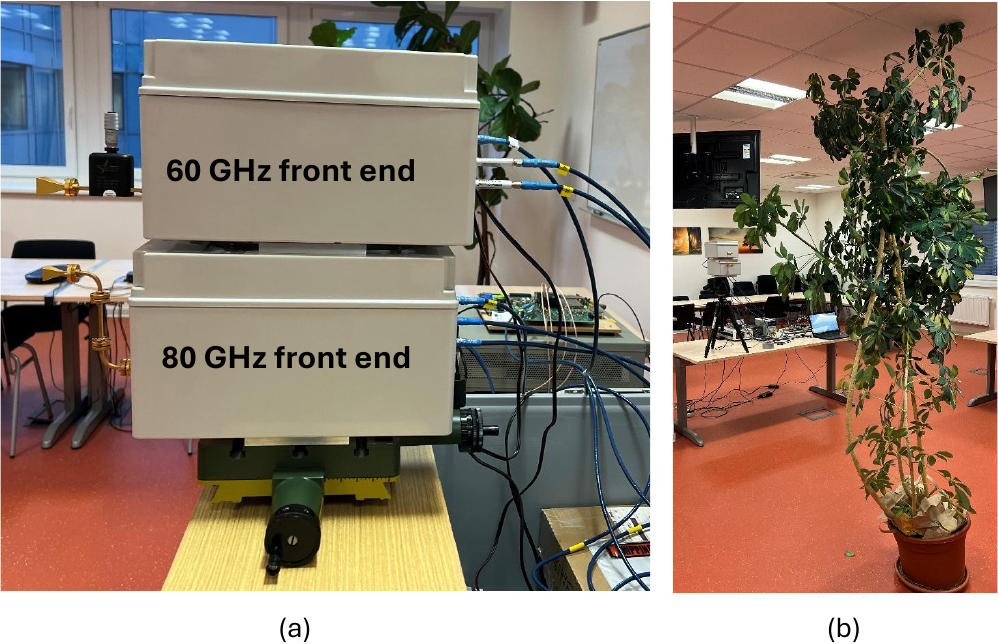}
    \caption{Photographs of the measurement scenario. (a) Detail of the 60\,GHz and 80\,GHz front ends enclosed in boxes, with horn antennas, all mounted on an XY tables. (b) The plant and the \ac{RX} side of the setup.}
    \label{fig:photo_of_env}
\end{figure}

The measurements were conducted in an empty conference room. The block diagram of the setup is shown in Fig. \ref{fig:diagram}. The \ac{TX} and \ac{RX} were positioned facing each other, separated by a distance of 4.5 m. The antennas were mounted vertically aligned, one above the other, at \ac{H} of 1.5\,m for the 80\,GHz band and 1.6\,m for the 60\,GHz band.
A potted plant, representing a simplified model of vegetation, was placed at the midpoint between the \ac{TX} and \ac{RX}, with its center positioned 2.4\,m from the \ac{TX}. The diameter of the plant at a height of 1.5\,m above the ground is approximately 0.8 m, and its total height is 2.8 m. A photograph of the measurement setup is provided in Fig.~\ref{fig:photo_of_env}, while Tab. \ref{tab:scenario} summarizes 
the relative distances between the \ac{TX}, \ac{RX}, and the plant. The plant was deliberately positioned to obstruct the direct \ac{LOS} path.
Both the transmitter and receiver were mounted on XY tables, allowing precise positioning in the x and y directions. For both frequency bands, the \ac{CIR} was measured for each incremental change in antenna position. The antenna displacement varied from fractions to multiple wavelengths, corresponding to the respective wavelengths at 60\,GHz and 80\,GHz. Based on the above, it is evident that the measurements were conducted in accordance with far-field conditions.

\begin{table}[h]
    \centering
    \caption{Distances between \ac{TX}, \ac{RX}, and the plant, as well as the heights (H) of the antennas above the ground.}
    \begin{tabular}{r||c|c|c|c}
        &RX-TX&TX-Plant&H (60\,GHz)&H (80\,GHz)\\
     \hline
      \hline
     Distance [m] & 4.75 & 2.4 & 1.6 & 1.5 \\
    \end{tabular}
    
    \label{tab:scenario}
\end{table}

\section{Measurement setup}\label{section_MS}
\begin{figure*}[htbp]
    \centering
    \includegraphics[width=1\textwidth]{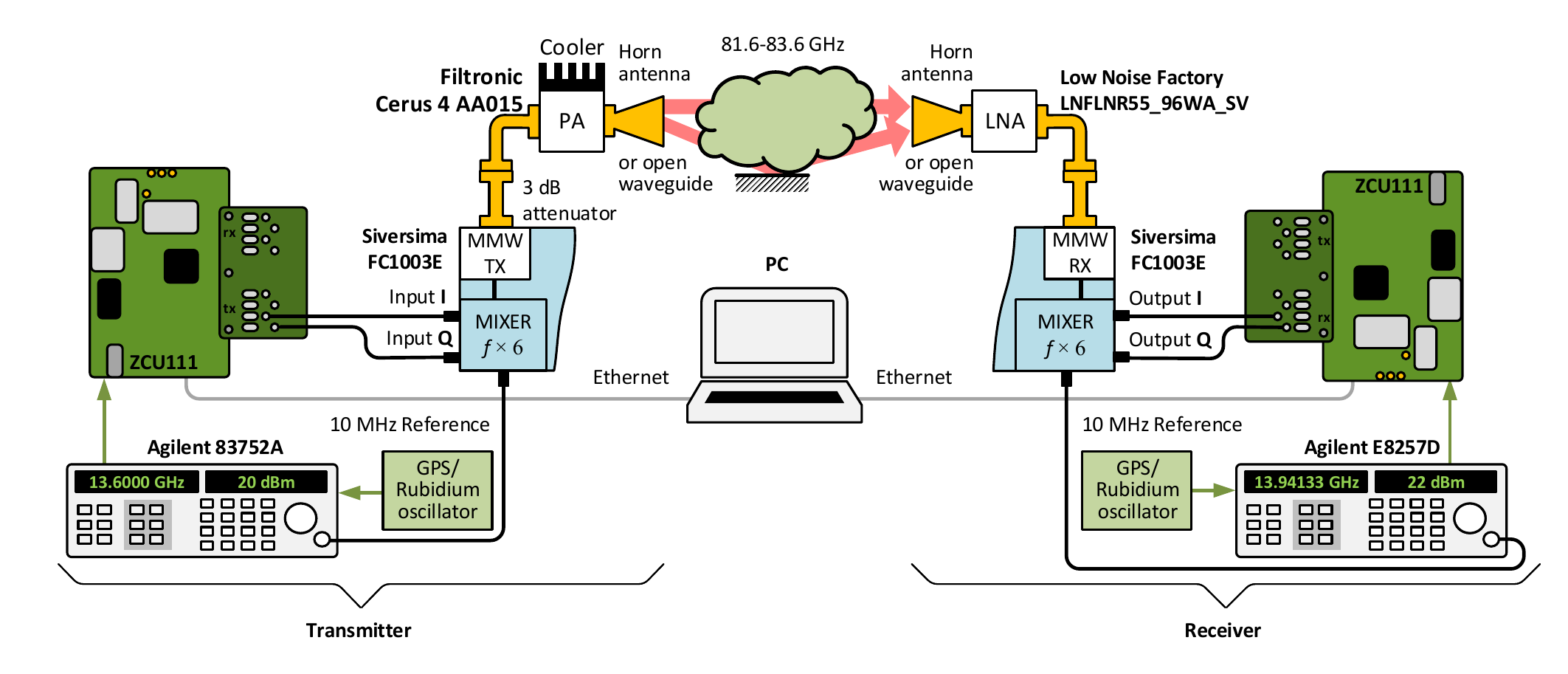}
    \caption{Measurement system schematic \cite{access_radek}}
    \label{fig:meas_schema}
\end{figure*}

The schematic representation of the measurement setup is shown in Fig.~\ref{fig:meas_schema}. The baseband transmission functionality is implemented using the Xilinx Zynq UltraScale+ RFSoC ZCU111 board. This platform generates the I and Q components of an intermediate frequency signal through high-speed DACs operating at 6.144\,GSPS. As the excitation signal, a \ac{FMCW} waveform with both rising and falling frequency slopes was selected to ensure a maximally uniform spectrum across the desired bandwidth.

The choice of an \ac{FMCW} waveform is primarily due to its robustness against non-linearities in RF hardware. With a bandwidth of $B=2.048\,\mathrm{GHz}$ and a sweep duration of $T=8\,\mu s$, the waveform enables rapid measurements with an update rate of $f_{\mathrm{meas}}=\frac{1}{T}=125\,\mathrm{kHz}$, while maintaining an acceptable \ac{SNR}. In static scenarios, further averaging can be applied to improve the signal quality.

To upconvert the signal to \ac{MMW} frequencies, a Sivers IMA up/down converter is used, model FC1005V/00 for 60\,GHz and FC1003E/03 for 80\,GHz. The local oscillator input is provided by an Agilent 83752A signal generator, offering high frequency stability and low phase noise. The RF signal is then amplified using a power amplifier, QuinStar QPW-50662330-C1 for 60\,GHz or Filtronic Cerus 4 AA015 for 80\,GHz, before being radiated through a horn antenna. A 10\,MHz GPS-disciplined oscillator ensures frequency synchronization throughout the system.

\begin{figure}[htbp]
    \centering
    \includegraphics[width=0.49\textwidth]{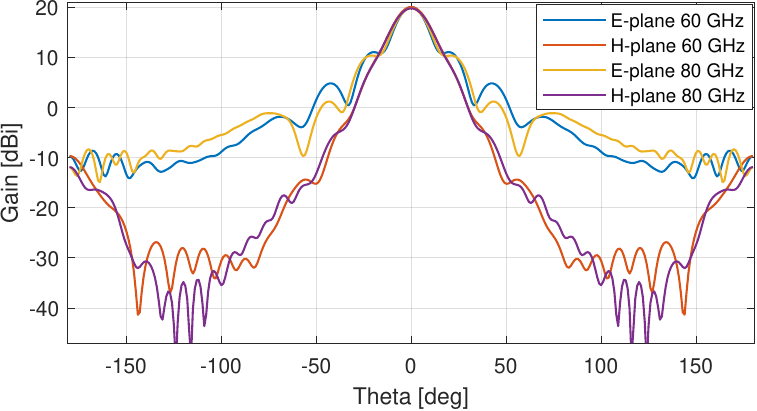}
    \caption{Simulated radiation pattern of horn antennas at 60\,GHz and 80\,GHz.}
    \label{fig:rad_pat_60_80}
\end{figure}

The radiation patterns of the 60\,GHz and 80\,GHz horn antennas are shown in Fig. \ref{fig:rad_pat_60_80}. Additional information about both antennas for these frequency bands is provided in Tab. \ref{tab:ant_parameters60_80}.

\begingroup
\setlength{\tabcolsep}{20pt} 
\begin{table}[h]
    \centering
    \caption{Parameters of horn antennas at 60\,GHz and 80\,GHz}
    \begin{tabular}{l||c|c}
        &60\,GHz & 80\,GHz\\
     \hline
      \hline
     Gain [dBi]&20& 20\\
     \hline
     HPBW E-plane [°]&14&16\\

    \end{tabular}
    
    \label{tab:ant_parameters60_80}
\end{table}
\endgroup

On the receiving side, a similar signal processing chain is used. After traversing the propagation environment, the signal is captured by a horn antenna and amplified using a low-noise amplifier, Quinstar QLW-50754530-I2 for 60,GHz or LNF-LNR55\_96WA\_SV by Low Noise Factory for 80\,GHz. The downconversion is carried out by a Sivers IMA FC1003V/01 (60\,GHz) or FC1003E/02 (80\,GHz) module, with the LO signal supplied by an Agilent E8257D generator. The I and Q components of the resulting intermediate frequency signal are digitized using the fast ADCs (4.096\,GSPS) of another ZCU111 board, and stored to an SSD for post-processing. Additional details on the testbed and calibration procedures are available in~\cite{access_radek}.

\section{Characterization of propagation properties}
\label{sec_char_CIR}

\subsection{Free space path loss at 60 and 80\,GHz}
There is a high attenuation at millimeter waves propagated in an open environment. The level of attenuation can be calculated according to (\ref{rce:FSPL})

\begin{equation}
\label{rce:FSPL}
\begin{split}
\mathrm{FSPL} & = \mathrm{20 \log_{10}}(d) + \mathrm{20 \log_{10}}(f) + \mathrm{20 \log_{10}}\left(\frac{4\pi}{c}\right), \\
\end{split}
\end{equation}

where $d$ represents the distance between the \ac{TX} and \ac{RX} antennas, $f$ is the center frequency, and $c$ is the speed of light. The FSPL attenuation is 81.07\,dB at 60\,GHz and 83.57\,dB at 80\,GHz band. In addition to loss of the path to the free space, interactions with oxygen ($O_2$), water or water vapor ($H_2O$) and other airborne molecules cause considerable attenuation of signal propagation in the \ac{MMW} band at particular carrier frequencies~\cite{graf_utlumu}. At carrier frequencies of about 60\,GHz, signals sent through the air are heavily absorbed, up to 15\,dB/km, due to the resonance frequency of ($O_2$) molecules. In contrast, the 80\,GHz band lies in the valley, with an additional attenuation of only 0.4\,dB/km. Over the 4.5\,m distance between the \ac{TX} and \ac{RX}, the additional attenuation is negligible but can play a significant role in real outdoor signal propagation.

\subsection{Signal forward scattering analysis}
\begin{figure}[htp]

\subfloat[]{%
  \includegraphics[clip,width=\columnwidth]{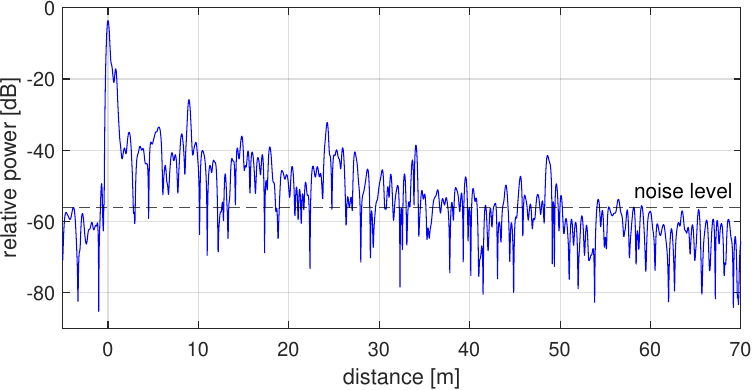}%
}

\subfloat[]{%
  \includegraphics[clip,width=\columnwidth]{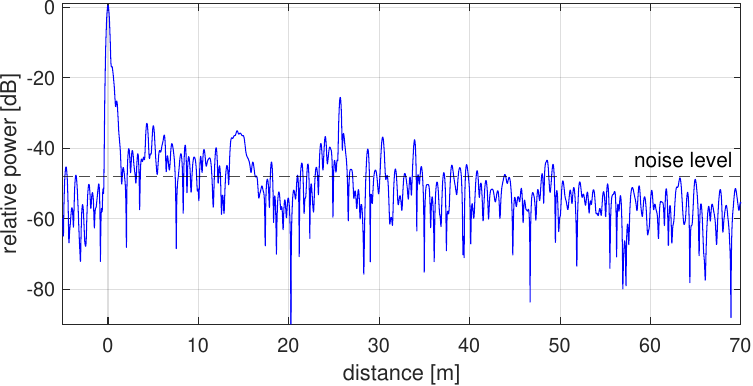}%
}

\caption{Measured \ac{PDP} for \ac{RX} position = 0, \ac{TX} position = 0. (a)~60\,GHz frequency band, (b) 80\,GHz frequency band.}
\label{fig:cir_60_80}
\end{figure}

This section examines the influence of plant foliage on signal propagation and analyzes the effect of micro-shifts in the \ac{TX} or \ac{RX} position on the received \ac{RP}.
We used our time domain channel sounder to estimate the \ac{CIR}~\cite{Hlawatsch2011}
\begin{equation}
\label{rce:CIR}
h_m(n)=\sum_{k=1}^{N} \alpha_k\mathrm{e}^{j2\pi f_D t} \delta(\tau-\tau_k),
\end{equation}

where $m$ is a measurement index and $N$ is the number of propagation paths. 
The variables $\alpha_k$ and $\tau_k$ corresponds to the gain and delay coefficient of the \mbox{$n$-th} multipath component while $\delta$ is the Dirac impulse and $f_D$ is the Doppler frequency. The \ac{CIR} as a function of distance is depicted in Fig. \ref{fig:cir_60_80} for \ac{RX} position~0 and \ac{TX} position~0, corresponding to the description of the positions in Fig.~\ref{fig:diagram}.


To analyze the influence of \ac{TX} or \ac{RX} antenna micro-shifts on the received power distribution within the \ac{MPC}, the average received \ac{RP} was calculated for three cases: \ac{LOS} (the first dominant component in the \ac{PDP}), \ac{NLOS} (all multipath components in the \ac{PDP} with the dominant \ac{LOS} component suppressed), and the combined \ac{LOS} + \ac{NLOS} components derived from the \ac{PDP}:


\begin{equation}
\label{rce:refl_pow}
RP = \frac{1}{N}\sum^N_{n=1}P(n),
\end{equation}
where $P(n)$ corresponds to the \ac{PDP}, determined as the average over several realizations according to:
\begin{equation}
\label{rce:E}
P(n)= \mathsf{E}\{|h(n)|^2\}.
\end{equation}

The \ac{RP} is computed from \ac{MPC} that are at least 6\,dB above the noise level, which is -56\,dB for the 60\,GHz band and -48\,dB for the 80\,GHz band.

\begin{figure}[htp]

\subfloat[]{%
  \includegraphics[clip,width=\columnwidth]{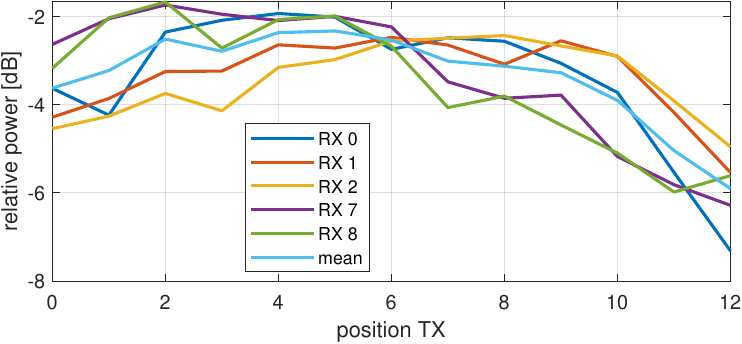}%
  \label{fig:pow_LOSa}
}

\subfloat[]{%
  \includegraphics[clip,width=\columnwidth]{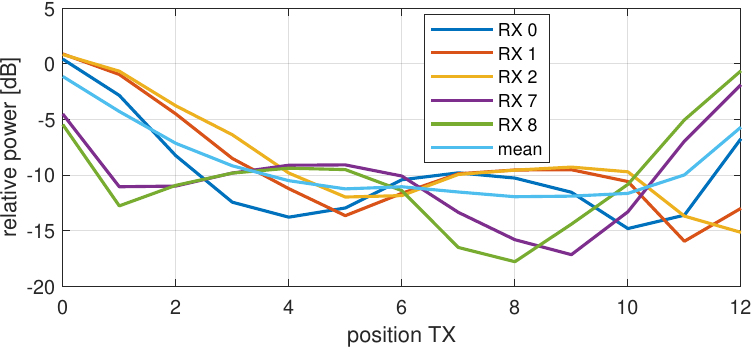}%
}

\caption{Relative power of the \ac{LOS} component in (a) the 60\,GHz band and (b) the 80\,GHz band.}
\label{fig:pow_LOS}
\end{figure}

The graphs depicting \ac{LOS} power for various \ac{RX} and \ac{TX} positions are shown in Fig. \ref{fig:pow_LOS}. Since both the \ac{TX} and \ac{RX} antennas were positioned 10 cm above each other, they were directed to different points on the plant. As a result, the electromagnetic waves experience different refraction and scattering, making it impossible to directly compare the shape of the power distribution curves at 60\,GHz and 80\,GHz. However, each frequency band can be statistically evaluated, allowing for a comparative analysis of the results.

As shown in Fig. \ref{fig:pow_LOSa}, the received power for the \ac{LOS} at 60\,GHz fluctuates between -2\,dB and -7\,dB across all \ac{TX} and \ac{RX} antenna positions. In the 80\,GHz, the fluctuation of \ac{LOS} relative power is more than three times greater compared to 60\,GHz band, the power dissipation is from 0\,dB to -18\,dB. This indicates that the 80\,GHz band is more sensitive to precise antenna positioning.
\begin{table*}[t]
    \centering
    \caption{Standard deviation of received relative power for different antennas positions}
    \label{tab:std_deviation}
    \renewcommand\arraystretch{1.2}  
    \begin{tabular}{c|c c c|c c c}
        \hline
        \multirow{2}{*}{RX Position} & 
        \multicolumn{3}{c|}{60\,GHz ($\sigma$, dB)} & 
        \multicolumn{3}{c}{80\,GHz ($\sigma$, dB)} \\
        \cline{2-7}
        & LOS+NLOS & LOS & NLOS & LOS+NLOS & LOS & NLOS  \\
        \hline
        0  & 0.8025  & 1.5099  & 0.7623  & 1.6221  & 4.3694  & 0.8487  \\
        1  & 0.4551  & 0.8631  & 0.5764  & 1.9548  & 4.6779  & 0.5882  \\
        2  & 0.5067  & 0.8296  & 0.5954  & 1.9707  & 4.6404  & 0.6208  \\
        7  & 0.6931  & 1.5146  & 0.5999  & 1.6157  & 4.2509  & 0.7024  \\
        8  & 0.5525  & 1.4114  & 0.6638  & 1.7391  & 4.7123  & 0.8747  \\
        \hline
    \end{tabular}
\end{table*}

\begin{figure}[htp]

\subfloat[]{%
  \includegraphics[clip,width=\columnwidth]{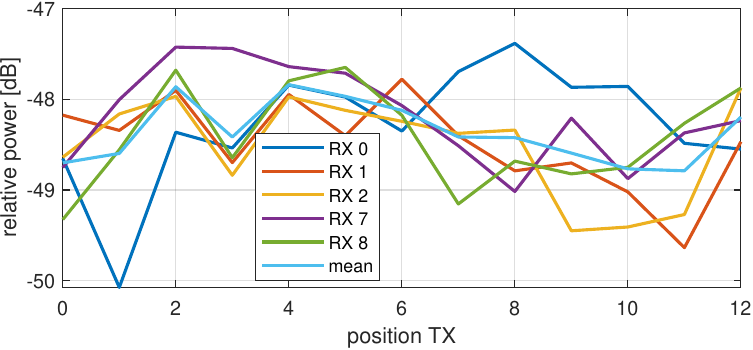}%
}

\subfloat[]{%
  \includegraphics[clip,width=\columnwidth]{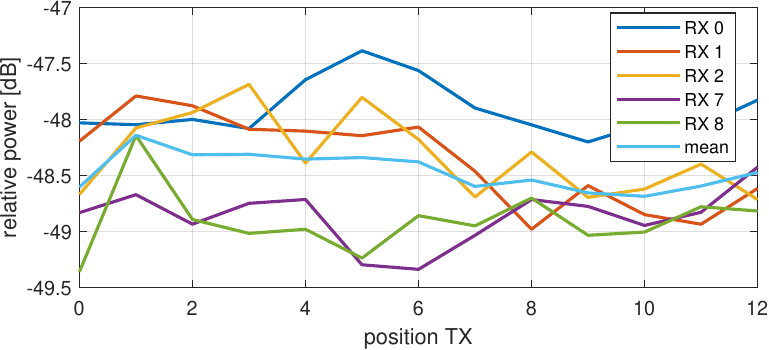}%
}

\caption{Relative power of the \ac{NLOS} component in (a) the 60\,GHz band and (b) the 80\,GHz band.}
\label{fig:pow_NLOS}
\end{figure}

In Fig. \ref{fig:pow_NLOS}, the graphs for the \ac{NLOS} \ac{MPC} are presented, with the dominant \ac{LOS} component suppressed. It is evident that the power is significantly lower compared to \ac{LOS}. The mean power is approximately -48\,dB for 60\,GHz and -48.5\,dB for 80\,GHz.

\begin{figure}[htp]

\subfloat[]{%
  \includegraphics[clip,width=\columnwidth]{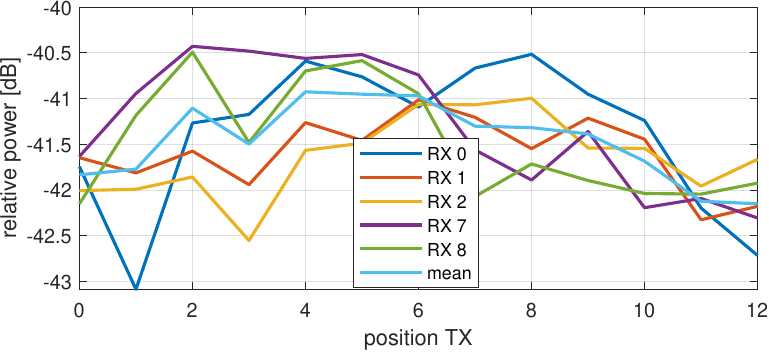}%
}

\subfloat[]{%
  \includegraphics[clip,width=\columnwidth]{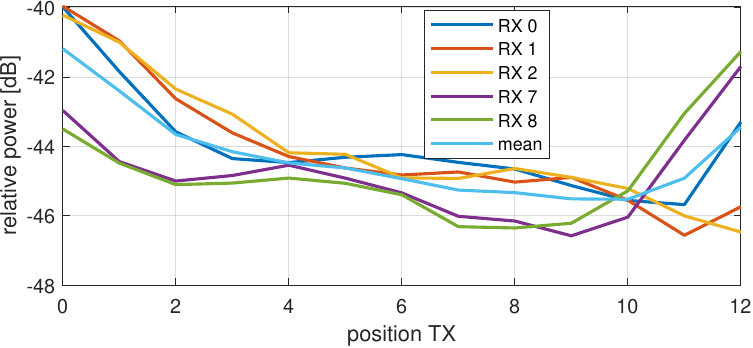}%
}

\caption{Relative power of the \ac{LOS} and \ac{MPC} component in (a) the 60\,GHz band and (b) the 80\,GHz band.}
\label{fig:pow_LOS_NLOS}
\end{figure}

The total received power, combining both the \ac{LOS} component and the \ac{NLOS} components, is presented in the graphs in Fig. \ref{fig:pow_LOS_NLOS} for each frequency band and various \ac{RX} and \ac{TX} positions. The plots clearly indicate that the shape of the curves follows that of the \ac{LOS} power profiles shown in Fig. \ref{fig:pow_LOS}, as the \ac{LOS} component is dominant and defines the overall shape. However, the total power is significantly lower, approximately 30\,dB less. A greater variance is observed in the 80\,GHz band, with fluctuations reaching up to 6\,dB, compared to the 60\,GHz band, where the variance is under 3\,dB.

Further comparison is provided in Tab. \ref{tab:std_deviation}, which evaluates the standard deviation for different \ac{RX} antenna positions in relation to the average value calculated from all measured \ac{TX} positions. The results are consistent with the previously shown graphs. For all \ac{LOS}/\ac{NLOS} combinations, the standard deviation is higher in the 80\,GHz band, which may be attributed to the shorter wavelength of the 80\,GHz signal compared to the 60\,GHz band.

\section{Conclusion}\label{section_C}
This paper investigated the effects of vegetation and small-scale antenna displacements on signal propagation at 60\,GHz and 80\,GHz through a controlled measurement campaign. A potted plant was placed in the direct \ac{LOS} path between the \ac{TX} and \ac{RX} to analyze its impact on attenuation and scattering. Additionally, the influence of micro-movements of the \ac{TX} and \ac{RX} antennas on the \ac{CIR} was examined.
The results indicate that the presence of vegetation introduces significant additional attenuation. Furthermore, the standard deviation of received power was higher at 80\,GHz, reaching up to 2\,dB, whereas at 60\,GHz, the variation remained below 1\,dB. This suggests that the 80\,GHz band is more sensitive to small-scale positional changes due to its shorter wavelength. The analysis of \ac{LOS} and \ac{NLOS} \ac{MPC} further confirms that vegetation contributes to increased signal fluctuations and scattering effects.



\section*{Acknowledgment}
The research described in this paper was financed by the Czech Science Foundation, Project No. 23-04304L, Multi-band prediction of millimeter-wave propagation effects for dynamic and fixed scenarios in rugged time varying environments and by the Internal Grant Agency of the Brno University of Technology under project no. FEKT-S-23-8191. The work of R. Shukla and A. Chandra is supported by the Chips-to-Startup (C2S) program no. EE-9/2/2021-R\&D-E from MeitY, GoI. The work of J. Kelner, J. Wojtuń and C. Ziółkowski was funded by the National Science Centre, Poland, under the OPUS-22 (LAP) call in the Weave program, as part of research project no. 2021/43/I/ST7/03294, acronym ‘MubaMilWave’.

\bibliographystyle{IEEEtran}
{\footnotesize
\bibliography{IEEEabrv,biblio_yousef.bib}

@book{Hlawatsch2011,
  author = {Franz Hlawatsch and Gerald Matz},
  title = {Wireless communications over rapidly time-varying channels},
  publisher = {Academic Press},
  address = {Burlington, MA},
  year = {2011},
  edition = {1st ed},
  isbn = {0123744830},
}

@INPROCEEDINGS{ITU_ISM_60,
  author={ITU},
  title={Radio Regulations, Section IV. Radio Stations and Systems – Article 1.15, definition: Industrial, scientific and medical (ISM) applications (of radio frequency energy)}
}

@ARTICLE{wrc19,
  author={Wang, Haiming and Zhang, Peize and Li, Jing and You, Xiaohu},
  journal={China Communications}, 
  title={Radio propagation and wireless coverage of LSAA-based 5G millimeter-wave mobile communication systems}, 
  year={2019},
  volume={16},
  number={5},
  pages={1-18},
  doi={10.23919/j.cc.2019.05.001}}

@ARTICLE{access_radek,
  author={Zavorka, Radek and Mikulasek, Tomas and Vychodil, Josef and Blumenstein, Jiri and Chandra, Aniruddha and Hammoud, Hussein and Kelner, Jan M. and Ziółkowski, Cezary and Zemen, Thomas and Mecklenbräuker, Christoph and Prokes, Ales},
  journal={IEEE Access}, 
  title={Characterizing the 80 GHz Channel in Static Scenarios: Diffuse Reflection, Scattering, and Transmission Through Trees Under Varying Weather Conditions}, 
  year={2024},
  volume={12},
  number={},
  pages={144738-144749},
  doi={10.1109/ACCESS.2024.3472003}}

@article{mpc,
author = {Zhou, Xin and Zhong, Zhangdui and Bian, Xin and He, Ruisi and Sun, Ruoyu and Guan, Ke and Liu, Ke},
title = {Indoor wideband channel measurements and analysis at 11 and 14 GHz},
journal = {IET Microwaves, Antennas \& Propagation},
volume = {11},
number = {10},
pages = {1393-1400},
year = {2017}
}

@INPROCEEDINGS{canyon,
  author={Eller, Lukas and Svoboda, Philipp and Rupp, Markus},
  booktitle={2021 IEEE 93rd Vehicular Technology Conference (VTC2021-Spring)}, 
  title={Propagation-aware Gaussian Process Regression for Signal-Strength Interpolation along Street Canyons}, 
  year={2021},
  volume={},
  number={},
  pages={1-7},
  doi={10.1109/VTC2021-Spring51267.2021.9448812}}

@ARTICLE{canyon2,
  author={Di Simone, Alessio and Iodice, Antonio},
  journal={IEEE Transactions on Antennas and Propagation}, 
  title={Modal Expansion Approach for Electromagnetic Propagation in Street Canyons}, 
  year={2019},
  volume={67},
  number={4},
  pages={2103-2117},
  doi={10.1109/TAP.2018.2883683}}

@ARTICLE{vege_brazil,
  author={Ramos, Glaucio L. and Leonor, Nuno R. and da Silva Mello, Luiz and Moreira, Fernando J. S. and Rego, Cassio G. and Goncalves, Sandro T. M. and Caldeirinha, Rafael F. S.},
  journal={IEEE Access}, 
  title={Polarimetric Vegetation Propagation Measurements on a Brazilian University Campus Scenario}, 
  year={2024},
  volume={12},
  number={},
  pages={181498-181508},
  doi={10.1109/ACCESS.2024.3507535}}

@INPROCEEDINGS{vege_china,
  author={Zhang, Peize and Li, Jing and Wang, Haiming and Hong, Wei},
  booktitle={2019 IEEE International Symposium on Antennas and Propagation and USNC-URSI Radio Science Meeting}, 
  title={Measurement-Based Propagation Characteristics at 28 GHz and 39 GHz in Suburban Environment}, 
  year={2019},
  volume={},
  number={},
  pages={2121-2122},
  doi={10.1109/APUSNCURSINRSM.2019.8889362}}

@article{graf_utlumu,
  author    = {Harsh Tataria and Katsuyuki Haneda and Andreas F. Molisch and others},
  title     = {Standardization of Propagation Models for Terrestrial Cellular Systems: A Historical Perspective},
  journal   = {International Journal of Wireless Information Networks},
  volume    = {28},
  pages     = {20--44},
  year      = {2021},
  doi       = {10.1007/s10776-020-00500-9}
}

@INPROCEEDINGS{foliage1,
  author={Rappaport, Theodore S. and Deng, Sijia},
  booktitle={IEEE International Conference on Communication Workshop (ICCW)}, 
  title={73\,{GHz} wideband millimeter-wave foliage and ground reflection measurements and models}, 
  year={London, UK, June 2015},
  volume={},
  number={},
  pages={1238-1243},
  doi={10.1109/ICCW.2015.7247347}}

@article{leaf_water,
author = {Browne, Marvin and Yardimci, Nezih Tolga and Scoffoni, Christine and Jarrahi, Mona and Sack, Lawren},
title = {Prediction of leaf water potential and relative water content using terahertz radiation spectroscopy},
journal = {Plant Direct},
volume = {4},
number = {4},
pages = {e00197},
doi = {https://doi.org/10.1002/pld3.197},
url = {https://onlinelibrary.wiley.com/doi/abs/10.1002/pld3.197},
eprint = {https://onlinelibrary.wiley.com/doi/pdf/10.1002/pld3.197},
year = {2020}
}

@INPROCEEDINGS{chavero,
  author={Chavero, M. and Polo, V. and Ramos, F. and Marti, J.},
  booktitle={IEEE MTT-S International Microwave Symposium Digest}, 
  title={Impact of vegetation on the performance of 28\,{GHz} {LMDS} transmission}, 
  year={Anaheim, CA, USA, June 1999},
  volume={3},
  number={},
  pages={1063-1066},
  doi={10.1109/MWSYM.1999.779571}}

@INPROCEEDINGS{absorb,
  author={Yang, Hang and Mou, Haifeng and Sun, Chengnan and Guo, Zhenyang and Liu, Xichen and Ding, Siran and Zou, Xianbing and Gao, Xiang},
  booktitle={3rd International Conference on Geology, Mapping and Remote Sensing (ICGMRS)}, 
  title={E-Band Propagation Measurements and Initial Analysis for Long-Range Communication Over Sea}, 
  year={Zhoushan, China, April 2022},
  doi={10.1109/ICGMRS55602.2022.9849392}}

@INPROCEEDINGS{v2v,
  author={Kucharski, Maciej and Kissinger, Dietmar and Ng, Herman Jalli},
  booktitle={IEEE 18th Topical Meeting on Silicon Monolithic Integrated Circuits in RF Systems (SiRF)}, 
  title={A universal monolithic E-band transceiver for automotive radar applications and V2V communication}, 
  year={Anaheim, CA, USA, January 2018},

  doi={10.1109/SIRF.2018.8304216}}
}


\end{document}